\documentclass[aps,prb,superscriptaddress,twocolumn,floatfix,showpacs,amsmath]{revtex4-1}
\usepackage{graphicx}
\graphicspath{ {./images/} }
\usepackage{floatrow}
\usepackage{latexsym}
\usepackage[makeroom]{cancel}
\usepackage{dcolumn}
\usepackage{epstopdf}
\usepackage{float}
\floatstyle{plaintop}
\restylefloat{table}

\usepackage{color}
\usepackage{amsmath,latexsym,psfrag}
\usepackage{array}
\usepackage{dcolumn}
\usepackage{bm}
\usepackage{epstopdf}
\usepackage{soul}  
\usepackage[
colorlinks=true,
urlcolor=blue,
linkcolor=blue,
citecolor=blue]{hyperref}
\begin{document}
\begin{titlepage}
\title {The Impact of Nitrogen Doping on Structural and Electronic Properties of Titanium Sesquioxide, Ti$_{2}$O$_{3}$ : An \textit{ab-initio} Study}

\author{Lynet Allan}
\email{lynetamondi3@gmail.com}
\affiliation{ Department of Physics, School of Physical Sciences, University of Nairobi, P.O.Box 30197-00100 Nairobi Kenya.}

\author{George O Amolo}
\email{georgeamollo862@gmail.com}

\affiliation{ Materials Modeling Group,
School of Physics and Earth Sciences,
The Technical University of Kenya,
52428-00200, Nairobi, Kenya.}

\author{Julius Mwabora}
\affiliation{ Department of Physics, School of Physical Sciences, University of Nairobi, P.O.Box 30197-00100 Nairobi Kenya.} 

\author{ Silas Mureramanzi}
\affiliation{ Department of Physics, School of Physical Sciences, University of Nairobi, P.O.Box 30197-00100 Nairobi Kenya.}

\date{\today}

\begin{abstract}

Titanium based oxides are abundant, chemically stable, non-toxic, and highly versatile materials, with applications ranging from photovoltaics to catalysis. For rutile and anatase phases of TiO$_{2}$, the bandgap ranges from 3.0–3.2 eV, which are too large to absorb in the visible range (400 nm - 700 nm), resulting in poor photo-catalytic efficiency. Nitrogen doping into TiO$_{2}$ has been able to narrow it's  bandgap, forming an absorption tail in the visible-light region. However, TiO$_{2}$ has limits to which it can be doped, suggesting investigations of the oxygen-deficient Ti$_{2}$O$_{3}$. Using the state-of-the-art Density Functional Theory (DFT) as implemented in the Quantum ESPRESSO package, we report on the structural and electronic properties of corundum-type Ti$_{2}$N$_{2}$O (an example Ti$_{n}$N$_{2}$O$_{2n-3}$ compound with n=2). The most stable sample of the oxynitride (Ti$_2$N$_2$O-P1), has a bandgap of 2.2 eV, which is clearly near the middle of visible light, and has no in-gap states, suggesting that they are more efficient visible-light-driven materials for photocatalytic applications compared to TiO$_{2}$, TiO$_{2}$: N and Ti$_{2}$O$_{3}$.

\end{abstract}

\maketitle

\end{titlepage}

\section{Introduction}
Due to increasing global energy demands and the harmful environmental pollution caused by the combustion of fossil fuels, the quest for sustainable alternative energy resources has become a major area of research\cite{shaffer2019global}. Ti-based oxides have been intensively investigated as photo-catalysts due to their wide applications in environmental remediation and solar energy conversion \cite{daghrir2013modified,agustina2005review}. Despite having wide bandgaps of 3.0 eV and 3.2 eV, respectively, rutile and anatase phases of titanium dioxide (TiO$_{2}$) still attracts considerable attention \cite{aoki2019insulating,agustina2005review,atambo2019electronic,carloelectronic,fujishima2000titanium,li2016investigation,wu2013first,shirai2018water}. To improve the solar application efficiency of Ti-based oxides, there has been an attempt to narrow their bandgap to absorb a major part of the solar spectrum \cite{khan2012first}. Doping is considered to be one of the promising methods of narrowing the bandgap of the Ti-based oxides \cite{guo2011highly}. 
\\Previous \textit{ab-initio} works have been reported on doped TiO$_{2}$ compounds.  V, Cr, Mn, Fe, and Ni have been doped into TiO$_{2}$ at the Ti sites\cite{inturi2014visible,umebayashi2002analysis}, such cationic doping leads to the localized d-states deep in the bandgap of TiO$_{2}$, which would act as the recombination centers for photoexcited electrons and holes, lowering photocatalytic activity. An anion doping in TiO$_{2}$ with N, C, S, and B has been equally explored \cite{morikawa2001band,fujishima2008tio2,sakthivel2003daylight,schneider2014understanding, morikawa2001band}, such resulted in the p-states near the valence band being similar to other deep donor levels in the semiconductor\cite{schneider2014understanding}. Owing to the  finite number of oxygen vacancies, TiO$_{2}$ has limits to which it can be doped\cite{shevlin2010electronic}.  In this work, we increase nitrogen concentrations in titanium oxide systems by considering  compounds having the composition Ti$_{n}$ O$_{2n-1}$, rather than TiO$_{2}$ with a finite number of oxygen vacancies. These systems can be viewed as ordered oxygen-deficient TiO$_{2}$ variants with a lot of oxygen vacancies which allows nitrogen substitutions, hence, should be able to accommodate far greater nitrogen concentrations than TiO$_{2}$.
 In search of novel photocatalytic materials that are stable, efficient, and capable of being visible light driven, Wu and his co-workers\cite{wu2013first}, using  high throughput screening method within density functional theory (DFT)\cite{Kohn-65}, discovered Ti$_{n}$N$_{2}$O$_{2n-3}$ compound for the \textit{n}=3 case (Ti$_{3}$N$_{2}$O$_{3}$), which they predicted to be synthesizable, suitable for visible light (or UV light) absorption, and energetically favorable to catalyze the water-splitting reaction.
\\Corundum Ti$_{2}$O$_{3}$, a Ti based oxide with nominal Ti$_{n}$ O$_{2n-1}$, \textit{n}=2 configuration together with its oxynitrides Ti$_{n}$N$_{2}$O$_{2n-3}$, \textit{n}=2, (Ti$_{2}$N$_{2}$O), has been much less investigated. Here, we explore the structural and electronic properties of Corundum Ti$_{2}$O$_{3}$ using Density Functional Theory (DFT) and Density Functional Theory with the inclusion of Hubbard \textit {U} correction {(DFT$\pmb{+}$U)} \cite{himmetoglu2014hubbard} studies.  In particular, we employ nitrogen substitutional doping at oxygen sites. The structural and electronic properties of the pristine and doped oxides, with and without the inclusion of Hubbard \textit{U}, are simulated and analyzed.  This study provides complementary theoretical data on structural and electronic properties of anionic doping of Ti$_{2}$O$_{3}$ with nitrogen. \\

\section{Computational details}
\begin{figure}
    \centering
    \includegraphics[width=0.7\textwidth]{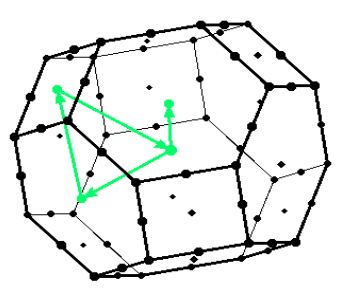}
    \caption{Brillouin zone integration for  Corundum Ti$_{2}$O$_{3}$ showing the origin at $\Gamma$}
    \label{fig:1}
\end{figure}

We have carried out simulations on trigonal Ti$_{2}$O$_{3}$ with a= b=c, $\alpha=\beta=\gamma \neq \ $90$^0$ within density functional theory (DFT)\cite{Kohn-65} using the generalized gradient approximation (GGA) as implemented in the Quantum ESPRESSO(QE) code\cite{Giannozzi_2009}. The electron-ion interactions are described by ultra-soft pseudo-potentials\cite{PAW-PPS} drawn from the 2.0.1 version of the library of Dal Corso et al,(2001)\cite{carloelectronic}. For the exchange-correlation interaction, the enhanced Perdew-Burke- Ernzerhof functional of the generalized gradient approximation, for solids (GGA-PBEsol) \cite{PBE-1996} are employed. This exchange-correlation is selected because it is computationally efficient and no adjustable parameter is required. An extra Hubbard \textit{U} term is included to account for the strong electron correlations, with  \textit{U} values tested in the range of 1 eV - 7 eV, applied only to the Ti-3d orbitals in  Ti$_{2}$O$_{3}$. The wave functions for valence electrons are expanded through a plane-wave basis set within the energy cut-off of 30 Ry. The atomic coordinates are relaxed until the forces are less than 0.01 eV/Å. The crystal structures are then viewed using the X-window Crystalline and Molecular Structure Visualization (XCrysDen)\cite{kokalj2001xcrysden} program where the lattice parameters \textit{a} and \textit{c} for the pristine and doped structures with and without the inclusion of the hubbard \textit{U}, are determined. A Monkhorst-Pack mesh sampling~\cite{Monkhorst-76} equivalent to  $5 \times 5 \times 5$ are employed.For band structure calculations, explicit positions along the high symmetry axes are described on the path ${\Gamma}$$\rightarrow{Z}$$\rightarrow{F}$$\rightarrow$${\Gamma}$$\rightarrow{L}$\cite{setyawan2010high} shown in Fig \ref{fig:1}. 

\section{Results and discussions}

\subsection{Structural Properties: Lattice Parameters}

By fitting the energy volume relationship to the Murnaghan equation of state\cite{murnaghan1944compressibility}, the volume that yields the minimum energy value are obtained. From table \ref{tab:table 1} the calculated lattice parameters \textbf{\textit a} and \textbf{\textit c } (at U=0 eV) lie slightly above the  experimental data\cite{li2018titanium}. It is well known that GGA tends to overestimate the lattice parameters\cite{niu2011enhanced}. The slight deviations are attributed to the fact that simulations have not taken into account the existence of intrinsic and other defects occurring in actual situations. To check on the feasibility of the doped systems, the structural properties of pure and doped Ti$_{2}$O$_{3}$ are compared. After the doping process, the lattice constants of doped systems are found to be slightly deformed, showing lattice strain due to the difference in radius between the dopant N atom(s) and the substituted O atom(s). The optimized lattice parameters for undoped Ti$_{2}$O$_{3}$ are a=b=c 5.274 Å, while for doped systems are a=b=c 5.271 Å. This result is consistent with the experimental results\cite{li2018titanium}, as well as other theoretical studies\cite{aoki2019insulating}, demonstrating the accuracy and dependability of the structural optimization parameters and method used.
\\
To account for the strong electron correlations, an extra Hubbard-U term is added, with a U\textit{$_{eff}$} of 1 eV - 7 eV applied only to the Ti-3d orbitals. As seen in Table \ref{tab:table 1}, there are some slight changes in the lattice parameter. To be more exact, the \textbf{c/a} ratio is seen to be decreasing with increasing values of \textit{U} at some point. The inclusion of the \textit{U} values in this computation is likely to have resulted in changes in the lattice parameters due to the correction of hybridization between O-2p and Ti-3d orbitals\cite{persson2006improved}. As a result, the ideal value of U is determined by seeking a fair agreement with experimental data for both structural and electronic properties.

\begin{table}[h]
\begin{tabular}{llll} \hline\hline
\textbf{ U Values }\\\textbf{(eV)} & \multicolumn{2}{l}{ \textbf{Lattice Parameters}}  & {\textbf{Band gaps (eV)} } \\ &\textbf{a(Å)} & {\textbf{c/a } }             &   \\
            \hline\hline  
0   & 5.27       & 1.03         & No gap                  \\
1   & 5.22       & 1.03          & 0.11                  \\
2   & 5.24      & 1.02           & 0.49                 \\
3   & 5.25      & 1.02           & 0.87                 \\
4   & 5.19     & 1.01            & 1.36                  \\
5   & 5.16     & 1.00            &1.63                    \\
6   & 5.38     & 1.00            & 1.76                 \\
7   & 5.35     & 1.03            & 1.86               \\
\textbf{Expt}\cite{li2018titanium}    & 5.15                                               & 1.01  & 1.65                       \\ \hline\hline                   
\end{tabular}
\caption{Variation of  lattice parameters and calculated band gaps with increasing Hubbard \textit{U} Values for Corundum Ti$_{2}$O$_{3}$} 
\label{tab:table 1}
\end{table}

\subsection{Choice of Hubbard \textit{U} for Ti$_{2}$O$_{3}$}
\begin{figure}[H]
   \centering
    \includegraphics[width=0.9\textwidth]{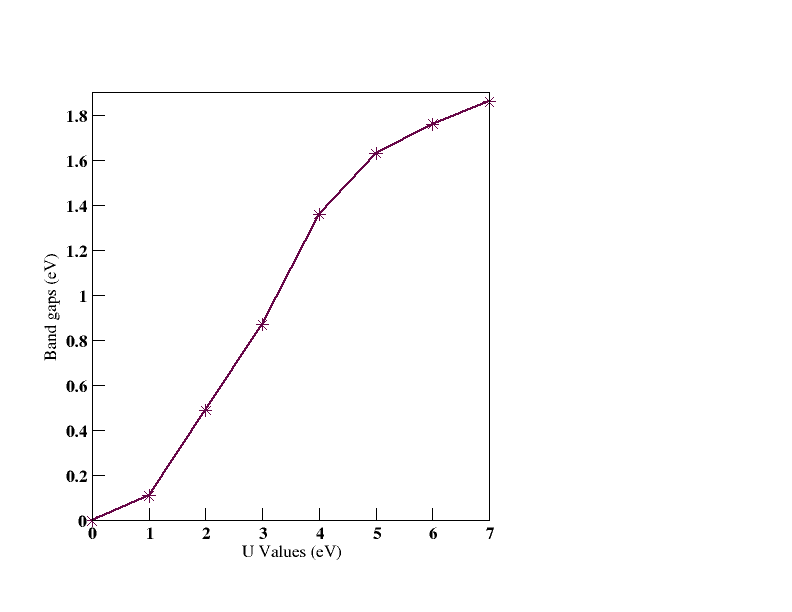}
    \caption{Variations of band gaps with increasing values of U}
    \label{Fig:2}
\end{figure}

Corundum Ti$_{2}$O$_{3}$, appeared metallic, as shown in Fig \ref{Fig:fig 10}\textbf{(a)}, this is consistent with other DFT findings \cite{aoki2019insulating}. Experimental studies by Li et al (2018)\cite{li2018titanium}, however, show that Corundum Ti$_{2}$O$_{3}$ has a bandgap of  1.65 eV. As expected, Kohn-Sham DFT calculations could not predict this small gap since it is known to underestimate the bandgaps of systems with \textit{d} and \textit{f} orbitals \cite{hu2011choice}. A moderated Hubbard potential,\textit{U}, applied to the calculations of electronic structures leads to more precise electronic band gaps. The \textit{U} values are optimized empirically by tuning the values of \textit{U} when seeking an agreement with the experimental results of the properties of the  Ti$_{2}$O$_{3}$\cite{hu2011choice}, as well as checking on the impact on lattice parameters as indicated in Table \ref{tab:table 1}. One thing that came out clear was that the value of the energy gap kept increasing in an almost linearly manner with the \textit{U} values as shown in Fig \ref{Fig:2}.  Corundum Ti$_{2}$O$_{3}$ opened a 1.638 eV band gap by DFT+U (\textit{U}=5 eV) studies, supporting the experimental data that this structure has semiconducting properties. The lattice parameter at \textit{U}=5 eV presented in Table \ref{tab:table 1} is also in good agreement with the experimental lattice parameter\cite{li2018titanium} of 5.15 eV. The value, \textit{U} = 5 eV, is therefore considered as the optimal value and inferred for improved prediction of the band gap of the corundum \-based Ti$_{2}$N$_{2}$O structures.

\subsection {Nitrogen substituted Ti$_{2}$O$_{3}$ (Ti$_{2}$N$_{2}$O)}

\begin{figure}[H]
    \centering
    \includegraphics[width=\textwidth]{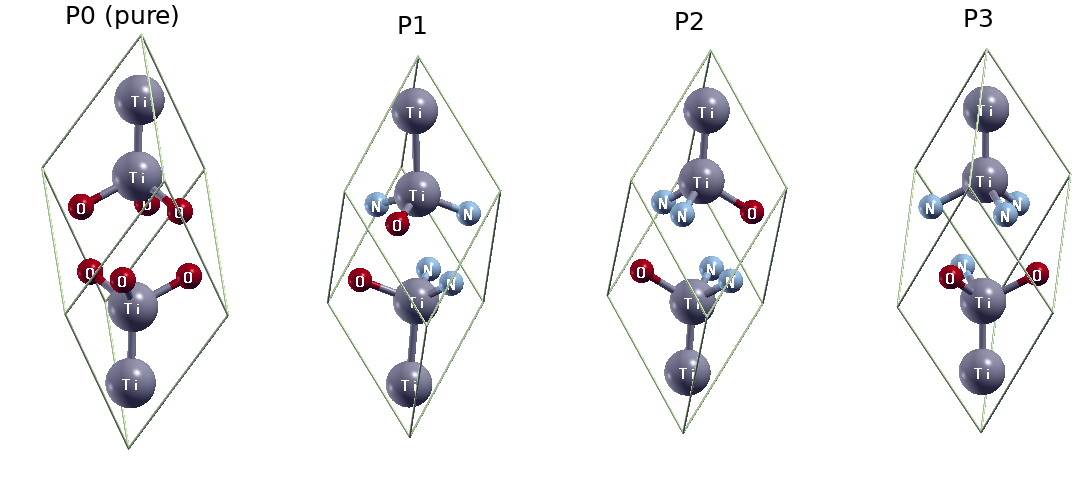}
    \caption{(Color online) (P0) Pure Corundum Ti$_{2}$O$_{3}$ [R3c (No. 167), trigonal],(P1)-(P3) Different orientations of Nitrogen substitutions in Ti$_{2}$O$_{3}$ structure. P2 has an inversion symmetry while P1 and P3 do not have.}
    \label{Fig:6}
\end{figure}

By substituting two of the ${2n-1}$ oxygen atoms of Ti$_{n}$O$_{2n-1}$ compounds with nitrogen, a series of Ti$_{2}$N$_{2}$O oxynitrides are modelled. Four out of six oxygen atoms are substituted with nitrogen to obtain  Ti$_{2}$N$_{2}$O compounds. Due to the symmetry of the  parent material, only three types (P1, P2, and P3) shown in Fig \ref{Fig:6} are considered.

\begin{figure}[H]
    \centering
    \includegraphics[width=0.9\textwidth]{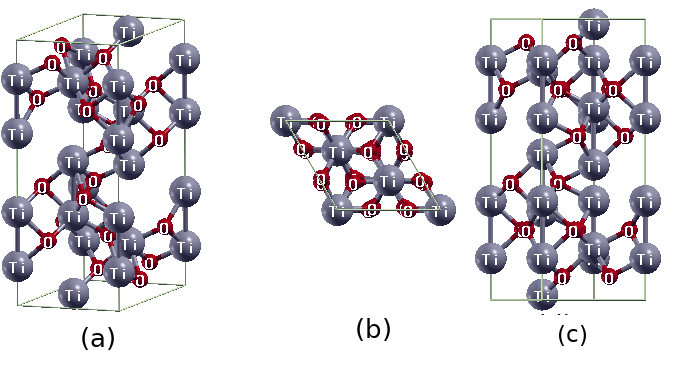}
    \caption{(Color online) Three different angles of projection of Ti$_{2}$O$_{3}$ hexagonal convectional cell (a) Bird's eye view (b) Top view, and (c) Projected in the direction of a+b.}
    \label{Fig:7}
\end{figure}

\begin{figure}[H]
    \centering
    \includegraphics[width=\textwidth]{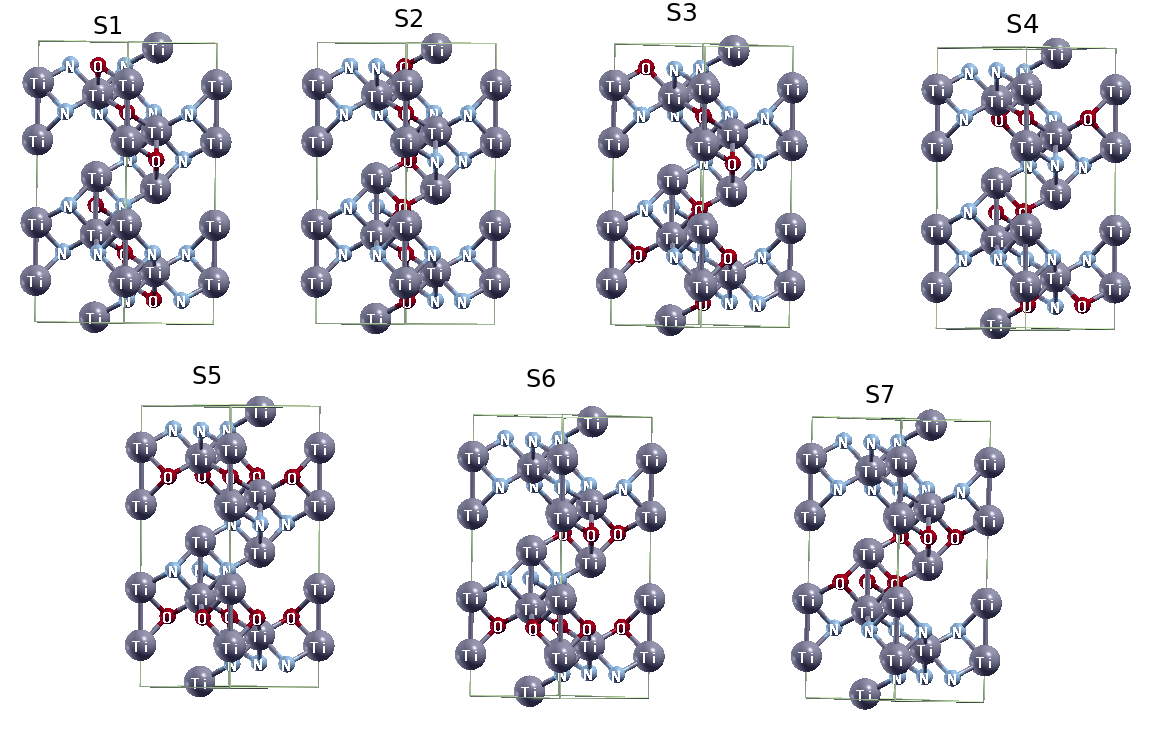}
    \caption{(Color online) Seven samples of Ti$_{2}$N$_{2}$O (S-structures)}
    \label{Fig:8}
\end{figure}

The other alternative is to model the oxynitrides in the hexagonal convectional cells of Ti$_{2}$O$_{3}$ containing 30 atoms (twelve titanium and eight oxygen atoms)\cite{aoki2019insulating}. This is as well considered and viewed at three different angles of projection in Fig \ref{Fig:7}.
Twelve out of eighteen oxygen atoms are replaced by nitrogen in seven different ways to form seven samples of Ti$_{2}$N$_{2}$O oxynitrides, projected in the a+b direction (for a clearer view), and arranged according to their stability as S1-S7, (S1 being the most stable sample) as shown in Fig \ref{Fig:8}. 

\begin{figure}[h]
    \centering
    \includegraphics[width=0.9\textwidth]{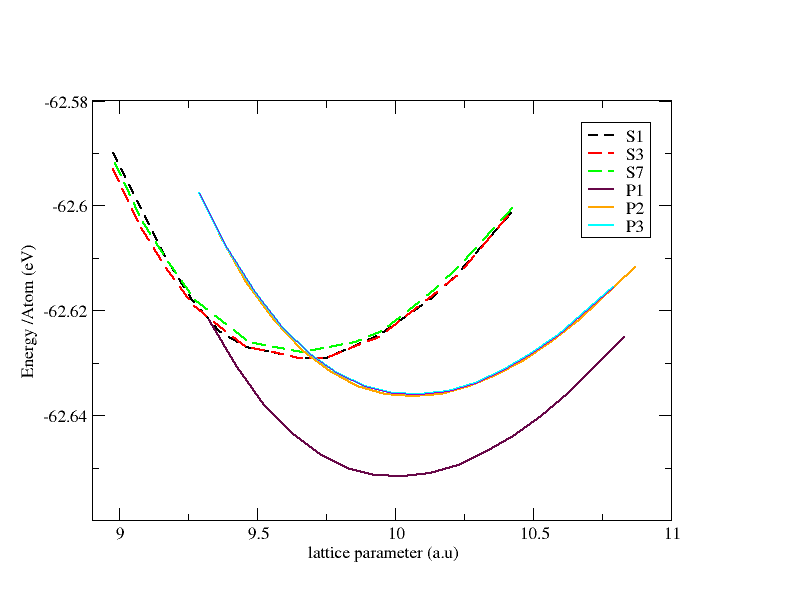}
    \caption{Plots for energy/cell (eV) against lattice parameter for both S- and P-structures.}
    \label{Fig:9}
\end{figure}

The stability of Ti$_{2}$N$_{2}$O (P and S) corundum based structures, are then compared as shown in Fig \ref{Fig:9}.

P-structures have lower formation energies than S-structures from Fig \ref{Fig:9}, hence more stable, with P1 being the most stable sample of all. This is well in agreement with the theory of Aoki et al.(2019)\cite{aoki2019insulating}, where GW studies were employed to study the structures. Therefore, the P structures are considered for further analysis of the band structures and projected density of state (PDOS) shown in Fig \ref{Fig:fig 10}.

\subsection{Band structures and Projected Density of States}

\begin{figure*}[h]
   \includegraphics[width=0.9\textwidth]{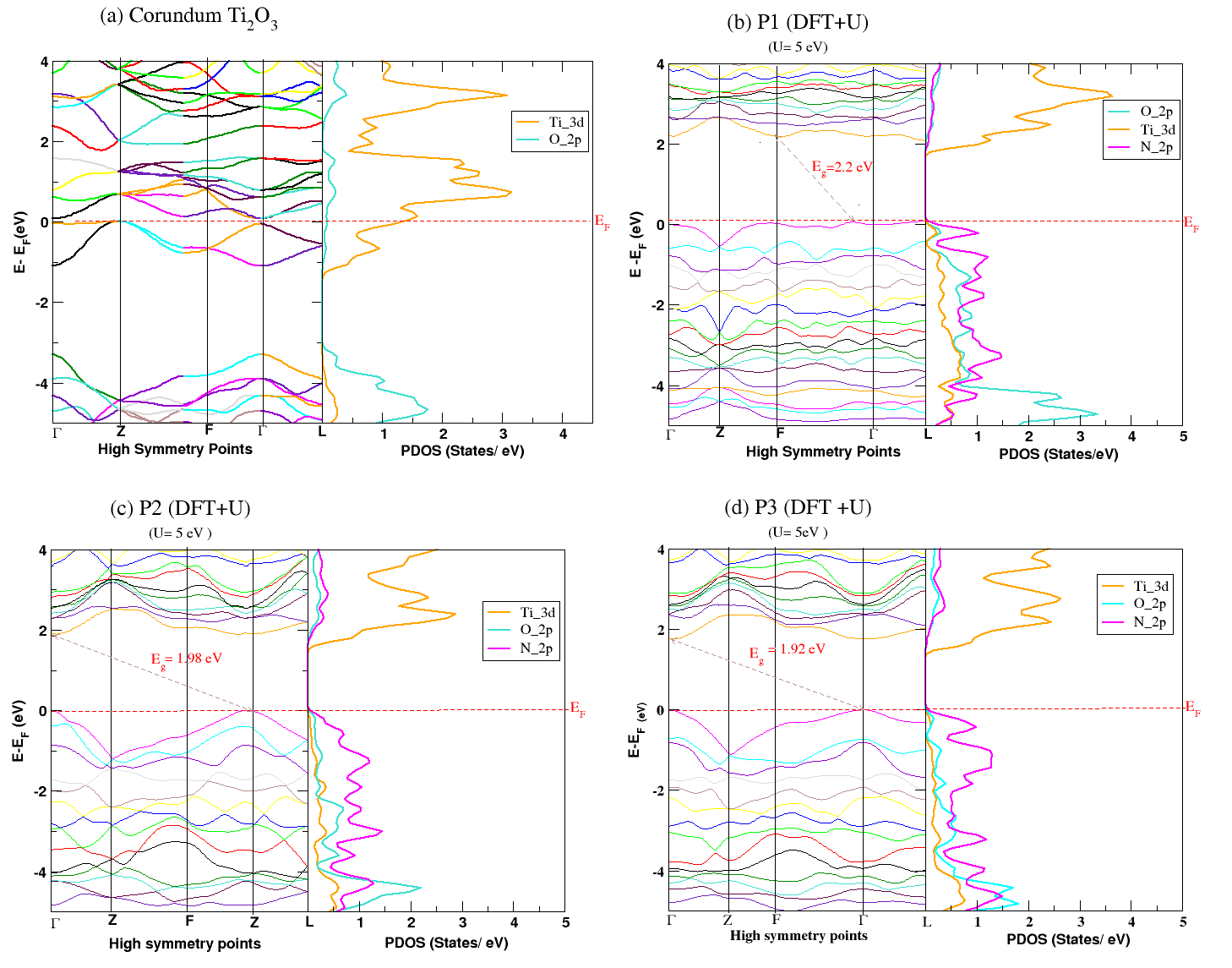}
    \caption{Band structures and Pdos for Corundum-based Ti$_{2}$N$_{2}$O\_P structures using DFT+U (U=5 eV) method.}
    \label{Fig:fig 10}
\end{figure*}

To explore the effect of doping on the electronic properties of Ti$_{2}$O$_{3}$ and clarify the origin of improved visible light photo activity, the band structures and projected density of states (PDOS) are calculated as given in Fig \ref{Fig:fig 10}.  For ease of comparison, the Fermi level (E$_{f}$) is set at zero. With  Ti-3d/ O-2p orbitals dominant in the entire band structure, Corundum Ti$_{2}$O$_{3}$ appeared metallic, as shown in Fig \ref{Fig:fig 10}\textbf{(a)}. 

Fig \ref{Fig:fig 10}\textbf{(b)-(d)} shows that nitrogen substitution into corundum Ti$_{2}$O$_{3}$ opens up a gap of 2.2 eV by DFT+U studies with U=5 eV in the P1 sample (most stable) while the P2 and P3 samples have band gaps of 1.98 eV and 1.92 eV, respectively. The electronic structure of Ti$_{2}$O$_{3}$  changes as the lattice parameters and deformation around the dopants changes, modifying their photocatalytic activity. In all three (P1, P2 and P3) samples, indirect bandgaps with bandgap character of Ti-3d/N-2p orbitals are observed. Some hybridization between N-2p and Ti-3d states is also observed in the valence band, although there are no in-gap states within the intrinsic gap, as there are in TiO$_2$ doped nitrogen \cite{persson2006improved}, thus the absorption threshold is unlikely to be affected. We have obtained a Ti-based oxide with a reduced band gap compared to the highly investigated TiO$_{2}$,  without necessarily creating the in-gap states which are known to serve as recombination centers for the photo-excited electrons and holes. This study suggests that, relative to pristine TiO$_{2}$, Ti$_{2}$O$_{3}$ and TiO$_{2}$: N compounds, Ti$_{2}$N$_{2}$O structures are more promising photo-catalytic/photo-voltaic materials. Table \ref{tab:table 5} offers a description of a comparison of the calculated and experimental bandgap for the oxides.

\begin{table*}[h]
\caption{Summary of the comparison of the calculated and experimental bandgap for the structures}
\label{tab:table 5}
\begin{tabular}{lllll} \\ \hline\hline
\textbf{MATERIAL}                                                                                                                            & \multicolumn{4}{l}{\textbf{ENERGY BAND GAP (eV)}}      \\
 & \textbf{\begin{tabular}[c]{@{}l@{}}THIS WORK\\   \\ (GGA)\end{tabular}}
 & \textbf{\begin{tabular}[c]{@{}l@{}}THEORETICAL\\   \\ (GGA)\end{tabular}}
 & \textbf{\begin{tabular}[c]{@{}l@{}}THIS WORK\\   \\ (DFT+U)\end{tabular}}                
 & \textbf{EXPERIMENTAL} \\ \hline\hline
                                           
Corundum Ti$_{2}$O$_{3}$  & No gap     & No gap {\cite{aoki2019insulating}} & 1.64(U=5 eV)        & 1.65{\cite{li2018titanium}}         \\
\begin{tabular}[c]{@{}l@{}}\textbf{Oxynitrides}\\   \\
Ti$_{2}$N$_{2}$O (P1)\\   \\
Ti$_{2}$N$_{2}$O (P2)\\   \\
Ti$_{2}$N$_{2}$O (P3)\end{tabular} &                                       &                                                          & \begin{tabular}[c]{@{}l@{}}2.20 (U=5 eV)\\  
                        \\ 1.98 (U=5 eV)\\  
                        \\ 1.92 (U=5 eV)\end{tabular} &   \\                  
\hline\hline
\end{tabular}

\end{table*}

\section{Conclusion}
Using the DFT and DFT+U approach, we have investigated the effect of nitrogen substitution on the geometry and electronic structure of corundum Ti$_{2}$O$_{3}$ in order to enhance its photo-catalytic performance. Significant structure information which may be useful in guiding experimental work as well as identifying appropriate applications for the material has been obtained. Structural and electronic properties of corundum Ti$_{2}$O$_{3}$ and the oxynitrides Ti$_{2}$N$_{2}$O have been analysed in this work. With regard to the impact of the Hubbard \textit{U} parameter on both structural and electronic properties, the best value of \textit{U} is found to be  5 eV. Even though these calculations assume defect-free structures, the lattice parameters and band gap obtained for this value of \textit{U} are in good agreement with other DFT results and experimental data. The oxygen deficient corundum Ti$_{2}$O$_{3}$, which showed metal-like properties opens up a bandgap of 1.638 eV by DFT+U (\textit{U}= 5 eV) studies, agreeing well with reported experimental data. To enhance photo-activity of Ti-based oxides, without creating mid-gap states, we aimed at reducing the band gaps of the oxides. This was achieved in Ti$_{2}$N$_{2}$O structures, which had a bandgap of 2.2 eV (563.56 nm), in the most stable sample\_P1, and had no in-gap states. In principle, Ti$_{2}$N$_{2}$O materials are found to be better for photo-catalytic/ photo-voltaic applications, compared to TiO$_{2}$, TiO$_{2}$:N and Ti$_{2}$O$_{3}$ materials.

\section{Acknowledgments}
This research was supported by the International Science Program (UON 500-661-127, PI- Prof. Aduda Bernard O.), through condensed matter thematic group of the Department of Physics, University of Nairobi. The Centre of High-Performance Computing (CHPC) through the project MATS862, Rosebank Cape Town Republic of South Africa is appreciated for providing access to the HPC cluster facility used in this research. Much appreciation to the Materials Modeling Group at the Technical University of Kenya for support.

\section{conflicts of interest}
There is no conflict of interest.

\bibliography{lyn}

\begin{thebibliography}{32}%
\makeatletter
\providecommand \@ifxundefined [1]{%
 \@ifx{#1\undefined}
}%
\providecommand \@ifnum [1]{%
 \ifnum #1\expandafter \@firstoftwo
 \else \expandafter \@secondoftwo
 \fi
}%
\providecommand \@ifx [1]{%
 \ifx #1\expandafter \@firstoftwo
 \else \expandafter \@secondoftwo
 \fi
}%
\providecommand \natexlab [1]{#1}%
\providecommand \enquote  [1]{``#1''}%
\providecommand \bibnamefont  [1]{#1}%
\providecommand \bibfnamefont [1]{#1}%
\providecommand \citenamefont [1]{#1}%
\providecommand \href@noop [0]{\@secondoftwo}%
\providecommand \href [0]{\begingroup \@sanitize@url \@href}%
\providecommand \@href[1]{\@@startlink{#1}\@@href}%
\providecommand \@@href[1]{\endgroup#1\@@endlink}%
\providecommand \@sanitize@url [0]{\catcode `\\12\catcode `\$12\catcode
  `\&12\catcode `\#12\catcode `\^12\catcode `\_12\catcode `\%12\relax}%
\providecommand \@@startlink[1]{}%
\providecommand \@@endlink[0]{}%
\providecommand \url  [0]{\begingroup\@sanitize@url \@url }%
\providecommand \@url [1]{\endgroup\@href {#1}{\urlprefix }}%
\providecommand \urlprefix  [0]{URL }%
\providecommand \Eprint [0]{\href }%
\providecommand \doibase [0]{http://dx.doi.org/}%
\providecommand \selectlanguage [0]{\@gobble}%
\providecommand \bibinfo  [0]{\@secondoftwo}%
\providecommand \bibfield  [0]{\@secondoftwo}%
\providecommand \translation [1]{[#1]}%
\providecommand \BibitemOpen [0]{}%
\providecommand \bibitemStop [0]{}%
\providecommand \bibitemNoStop [0]{.\EOS\space}%
\providecommand \EOS [0]{\spacefactor3000\relax}%
\providecommand \BibitemShut  [1]{\csname bibitem#1\endcsname}%
\let\auto@bib@innerbib\@empty
\bibitem [{\citenamefont {Shaffer}(2019)}]{shaffer2019global}%
  \BibitemOpen
  \bibfield  {author} {\bibinfo {author} {\bibfnamefont {B.}~\bibnamefont
  {Shaffer}},\ }\href@noop {} {\bibfield  {journal} {\bibinfo  {journal} {MRS
  Energy \& Sustainability}\ }\textbf {\bibinfo {volume} {6}} (\bibinfo {year}
  {2019})}\BibitemShut {NoStop}%
\bibitem [{\citenamefont {Daghrir}\ \emph {et~al.}(2013)\citenamefont
  {Daghrir}, \citenamefont {Drogui},\ and\ \citenamefont
  {Robert}}]{daghrir2013modified}%
  \BibitemOpen
  \bibfield  {author} {\bibinfo {author} {\bibfnamefont {R.}~\bibnamefont
  {Daghrir}}, \bibinfo {author} {\bibfnamefont {P.}~\bibnamefont {Drogui}}, \
  and\ \bibinfo {author} {\bibfnamefont {D.}~\bibnamefont {Robert}},\
  }\href@noop {} {\bibfield  {journal} {\bibinfo  {journal} {Industrial \&
  Engineering Chemistry Research}\ }\textbf {\bibinfo {volume} {52}},\ \bibinfo
  {pages} {3581} (\bibinfo {year} {2013})}\BibitemShut {NoStop}%
\bibitem [{\citenamefont {Agustina}\ \emph {et~al.}(2005)\citenamefont
  {Agustina}, \citenamefont {Ang},\ and\ \citenamefont
  {Vareek}}]{agustina2005review}%
  \BibitemOpen
  \bibfield  {author} {\bibinfo {author} {\bibfnamefont {T.~E.}\ \bibnamefont
  {Agustina}}, \bibinfo {author} {\bibfnamefont {H.~M.}\ \bibnamefont {Ang}}, \
  and\ \bibinfo {author} {\bibfnamefont {V.~K.}\ \bibnamefont {Vareek}},\
  }\href@noop {} {\bibfield  {journal} {\bibinfo  {journal} {Journal of
  Photochemistry and Photobiology C: Photochemistry Reviews}\ }\textbf
  {\bibinfo {volume} {6}},\ \bibinfo {pages} {264} (\bibinfo {year}
  {2005})}\BibitemShut {NoStop}%
\bibitem [{\citenamefont {Aoki}\ \emph {et~al.}(2019)\citenamefont {Aoki},
  \citenamefont {Sakurai}, \citenamefont {Coh}, \citenamefont {Chelikowsky},
  \citenamefont {Louie}, \citenamefont {Cohen},\ and\ \citenamefont
  {Saito}}]{aoki2019insulating}%
  \BibitemOpen
  \bibfield  {author} {\bibinfo {author} {\bibfnamefont {Y.}~\bibnamefont
  {Aoki}}, \bibinfo {author} {\bibfnamefont {M.}~\bibnamefont {Sakurai}},
  \bibinfo {author} {\bibfnamefont {S.}~\bibnamefont {Coh}}, \bibinfo {author}
  {\bibfnamefont {J.~R.}\ \bibnamefont {Chelikowsky}}, \bibinfo {author}
  {\bibfnamefont {S.~G.}\ \bibnamefont {Louie}}, \bibinfo {author}
  {\bibfnamefont {M.~L.}\ \bibnamefont {Cohen}}, \ and\ \bibinfo {author}
  {\bibfnamefont {S.}~\bibnamefont {Saito}},\ }\href@noop {} {\bibfield
  {journal} {\bibinfo  {journal} {Physical Review B}\ }\textbf {\bibinfo
  {volume} {99}},\ \bibinfo {pages} {075203} (\bibinfo {year}
  {2019})}\BibitemShut {NoStop}%
\bibitem [{\citenamefont {Atambo}\ \emph {et~al.}(2019)\citenamefont {Atambo},
  \citenamefont {Varsano}, \citenamefont {Ferretti}, \citenamefont {Ataei},
  \citenamefont {Caldas}, \citenamefont {Molinari},\ and\ \citenamefont
  {Selloni}}]{atambo2019electronic}%
  \BibitemOpen
  \bibfield  {author} {\bibinfo {author} {\bibfnamefont {M.~O.}\ \bibnamefont
  {Atambo}}, \bibinfo {author} {\bibfnamefont {D.}~\bibnamefont {Varsano}},
  \bibinfo {author} {\bibfnamefont {A.}~\bibnamefont {Ferretti}}, \bibinfo
  {author} {\bibfnamefont {S.~S.}\ \bibnamefont {Ataei}}, \bibinfo {author}
  {\bibfnamefont {M.~J.}\ \bibnamefont {Caldas}}, \bibinfo {author}
  {\bibfnamefont {E.}~\bibnamefont {Molinari}}, \ and\ \bibinfo {author}
  {\bibfnamefont {A.}~\bibnamefont {Selloni}},\ }\href@noop {} {\bibfield
  {journal} {\bibinfo  {journal} {Physical Review Materials}\ }\textbf
  {\bibinfo {volume} {3}},\ \bibinfo {pages} {045401} (\bibinfo {year}
  {2019})}\BibitemShut {NoStop}%
\bibitem [{\citenamefont {Carlo}()}]{carloelectronic}%
  \BibitemOpen
  \bibfield  {author} {\bibinfo {author} {\bibfnamefont {Q.~M.}\ \bibnamefont
  {Carlo}},\ }\href@noop {} {\ }\BibitemShut {NoStop}%
\bibitem [{\citenamefont {Fujishima}\ \emph {et~al.}(2000)\citenamefont
  {Fujishima}, \citenamefont {Rao},\ and\ \citenamefont
  {Tryk}}]{fujishima2000titanium}%
  \BibitemOpen
  \bibfield  {author} {\bibinfo {author} {\bibfnamefont {A.}~\bibnamefont
  {Fujishima}}, \bibinfo {author} {\bibfnamefont {T.~N.}\ \bibnamefont {Rao}},
  \ and\ \bibinfo {author} {\bibfnamefont {D.~A.}\ \bibnamefont {Tryk}},\
  }\href@noop {} {\bibfield  {journal} {\bibinfo  {journal} {Journal of
  photochemistry and photobiology C: Photochemistry reviews}\ }\textbf
  {\bibinfo {volume} {1}},\ \bibinfo {pages} {1} (\bibinfo {year}
  {2000})}\BibitemShut {NoStop}%
\bibitem [{\citenamefont {Li}(2016)}]{li2016investigation}%
  \BibitemOpen
  \bibfield  {author} {\bibinfo {author} {\bibfnamefont {Y.}~\bibnamefont
  {Li}},\ }\emph {\bibinfo {title} {Investigation of Titanium Sesquioxide
  Ti2O3: Synthesis and Physical Properties}},\ \href@noop {} {Ph.D. thesis}
  (\bibinfo {year} {2016})\BibitemShut {NoStop}%
\bibitem [{\citenamefont {Wu}\ \emph {et~al.}(2013)\citenamefont {Wu},
  \citenamefont {Lazic}, \citenamefont {Hautier}, \citenamefont {Persson},\
  and\ \citenamefont {Ceder}}]{wu2013first}%
  \BibitemOpen
  \bibfield  {author} {\bibinfo {author} {\bibfnamefont {Y.}~\bibnamefont
  {Wu}}, \bibinfo {author} {\bibfnamefont {P.}~\bibnamefont {Lazic}}, \bibinfo
  {author} {\bibfnamefont {G.}~\bibnamefont {Hautier}}, \bibinfo {author}
  {\bibfnamefont {K.}~\bibnamefont {Persson}}, \ and\ \bibinfo {author}
  {\bibfnamefont {G.}~\bibnamefont {Ceder}},\ }\href@noop {} {\bibfield
  {journal} {\bibinfo  {journal} {Energy \& environmental science}\ }\textbf
  {\bibinfo {volume} {6}},\ \bibinfo {pages} {157} (\bibinfo {year}
  {2013})}\BibitemShut {NoStop}%
\bibitem [{\citenamefont {Shirai}\ \emph {et~al.}(2018)\citenamefont {Shirai},
  \citenamefont {Fazio}, \citenamefont {Sugimoto}, \citenamefont {Selli},
  \citenamefont {Ferraro}, \citenamefont {Watanabe}, \citenamefont {Haruta},
  \citenamefont {Ohtani}, \citenamefont {Kurata}, \citenamefont {Di~Valentin}
  \emph {et~al.}}]{shirai2018water}%
  \BibitemOpen
  \bibfield  {author} {\bibinfo {author} {\bibfnamefont {K.}~\bibnamefont
  {Shirai}}, \bibinfo {author} {\bibfnamefont {G.}~\bibnamefont {Fazio}},
  \bibinfo {author} {\bibfnamefont {T.}~\bibnamefont {Sugimoto}}, \bibinfo
  {author} {\bibfnamefont {D.}~\bibnamefont {Selli}}, \bibinfo {author}
  {\bibfnamefont {L.}~\bibnamefont {Ferraro}}, \bibinfo {author} {\bibfnamefont
  {K.}~\bibnamefont {Watanabe}}, \bibinfo {author} {\bibfnamefont
  {M.}~\bibnamefont {Haruta}}, \bibinfo {author} {\bibfnamefont
  {B.}~\bibnamefont {Ohtani}}, \bibinfo {author} {\bibfnamefont
  {H.}~\bibnamefont {Kurata}}, \bibinfo {author} {\bibfnamefont
  {C.}~\bibnamefont {Di~Valentin}},  \emph {et~al.},\ }\href@noop {} {\bibfield
   {journal} {\bibinfo  {journal} {Journal of the American Chemical Society}\
  }\textbf {\bibinfo {volume} {140}},\ \bibinfo {pages} {1415} (\bibinfo {year}
  {2018})}\BibitemShut {NoStop}%
\bibitem [{\citenamefont {Khan}\ \emph {et~al.}(2012)\citenamefont {Khan},
  \citenamefont {Xu}, \citenamefont {Chen},\ and\ \citenamefont
  {Cao}}]{khan2012first}%
  \BibitemOpen
  \bibfield  {author} {\bibinfo {author} {\bibfnamefont {M.}~\bibnamefont
  {Khan}}, \bibinfo {author} {\bibfnamefont {J.}~\bibnamefont {Xu}}, \bibinfo
  {author} {\bibfnamefont {N.}~\bibnamefont {Chen}}, \ and\ \bibinfo {author}
  {\bibfnamefont {W.}~\bibnamefont {Cao}},\ }\href@noop {} {\bibfield
  {journal} {\bibinfo  {journal} {Journal of Alloys and Compounds}\ }\textbf
  {\bibinfo {volume} {513}},\ \bibinfo {pages} {539} (\bibinfo {year}
  {2012})}\BibitemShut {NoStop}%
\bibitem [{\citenamefont {Guo}\ \emph {et~al.}(2011)\citenamefont {Guo},
  \citenamefont {Wu}, \citenamefont {Chen}, \citenamefont {Boschloo},
  \citenamefont {Hagfeldt},\ and\ \citenamefont {Ma}}]{guo2011highly}%
  \BibitemOpen
  \bibfield  {author} {\bibinfo {author} {\bibfnamefont {W.}~\bibnamefont
  {Guo}}, \bibinfo {author} {\bibfnamefont {L.}~\bibnamefont {Wu}}, \bibinfo
  {author} {\bibfnamefont {Z.}~\bibnamefont {Chen}}, \bibinfo {author}
  {\bibfnamefont {G.}~\bibnamefont {Boschloo}}, \bibinfo {author}
  {\bibfnamefont {A.}~\bibnamefont {Hagfeldt}}, \ and\ \bibinfo {author}
  {\bibfnamefont {T.}~\bibnamefont {Ma}},\ }\href@noop {} {\bibfield  {journal}
  {\bibinfo  {journal} {Journal of Photochemistry and Photobiology A:
  Chemistry}\ }\textbf {\bibinfo {volume} {219}},\ \bibinfo {pages} {180}
  (\bibinfo {year} {2011})}\BibitemShut {NoStop}%
\bibitem [{\citenamefont {Inturi}\ \emph {et~al.}(2014)\citenamefont {Inturi},
  \citenamefont {Boningari}, \citenamefont {Suidan},\ and\ \citenamefont
  {Smirniotis}}]{inturi2014visible}%
  \BibitemOpen
  \bibfield  {author} {\bibinfo {author} {\bibfnamefont {S.~N.~R.}\
  \bibnamefont {Inturi}}, \bibinfo {author} {\bibfnamefont {T.}~\bibnamefont
  {Boningari}}, \bibinfo {author} {\bibfnamefont {M.}~\bibnamefont {Suidan}}, \
  and\ \bibinfo {author} {\bibfnamefont {P.~G.}\ \bibnamefont {Smirniotis}},\
  }\href@noop {} {\bibfield  {journal} {\bibinfo  {journal} {Applied Catalysis
  B: Environmental}\ }\textbf {\bibinfo {volume} {144}},\ \bibinfo {pages}
  {333} (\bibinfo {year} {2014})}\BibitemShut {NoStop}%
\bibitem [{\citenamefont {Umebayashi}\ \emph {et~al.}(2002)\citenamefont
  {Umebayashi}, \citenamefont {Yamaki}, \citenamefont {Itoh},\ and\
  \citenamefont {Asai}}]{umebayashi2002analysis}%
  \BibitemOpen
  \bibfield  {author} {\bibinfo {author} {\bibfnamefont {T.}~\bibnamefont
  {Umebayashi}}, \bibinfo {author} {\bibfnamefont {T.}~\bibnamefont {Yamaki}},
  \bibinfo {author} {\bibfnamefont {H.}~\bibnamefont {Itoh}}, \ and\ \bibinfo
  {author} {\bibfnamefont {K.}~\bibnamefont {Asai}},\ }\href@noop {} {\bibfield
   {journal} {\bibinfo  {journal} {Journal of Physics and Chemistry of Solids}\
  }\textbf {\bibinfo {volume} {63}},\ \bibinfo {pages} {1909} (\bibinfo {year}
  {2002})}\BibitemShut {NoStop}%
\bibitem [{\citenamefont {Morikawa}\ \emph {et~al.}(2001)\citenamefont
  {Morikawa}, \citenamefont {Asahi}, \citenamefont {Ohwaki}, \citenamefont
  {Aoki},\ and\ \citenamefont {Taga}}]{morikawa2001band}%
  \BibitemOpen
  \bibfield  {author} {\bibinfo {author} {\bibfnamefont {T.}~\bibnamefont
  {Morikawa}}, \bibinfo {author} {\bibfnamefont {R.}~\bibnamefont {Asahi}},
  \bibinfo {author} {\bibfnamefont {T.}~\bibnamefont {Ohwaki}}, \bibinfo
  {author} {\bibfnamefont {K.}~\bibnamefont {Aoki}}, \ and\ \bibinfo {author}
  {\bibfnamefont {Y.}~\bibnamefont {Taga}},\ }\href@noop {} {\bibfield
  {journal} {\bibinfo  {journal} {Japanese Journal of Applied Physics}\
  }\textbf {\bibinfo {volume} {40}},\ \bibinfo {pages} {L561} (\bibinfo {year}
  {2001})}\BibitemShut {NoStop}%
\bibitem [{\citenamefont {Fujishima}\ \emph {et~al.}(2008)\citenamefont
  {Fujishima}, \citenamefont {Zhang},\ and\ \citenamefont
  {Tryk}}]{fujishima2008tio2}%
  \BibitemOpen
  \bibfield  {author} {\bibinfo {author} {\bibfnamefont {A.}~\bibnamefont
  {Fujishima}}, \bibinfo {author} {\bibfnamefont {X.}~\bibnamefont {Zhang}}, \
  and\ \bibinfo {author} {\bibfnamefont {D.~A.}\ \bibnamefont {Tryk}},\
  }\href@noop {} {\bibfield  {journal} {\bibinfo  {journal} {Surface science
  reports}\ }\textbf {\bibinfo {volume} {63}},\ \bibinfo {pages} {515}
  (\bibinfo {year} {2008})}\BibitemShut {NoStop}%
\bibitem [{\citenamefont {Sakthivel}\ and\ \citenamefont
  {Kisch}(2003)}]{sakthivel2003daylight}%
  \BibitemOpen
  \bibfield  {author} {\bibinfo {author} {\bibfnamefont {S.}~\bibnamefont
  {Sakthivel}}\ and\ \bibinfo {author} {\bibfnamefont {H.}~\bibnamefont
  {Kisch}},\ }\href@noop {} {\bibfield  {journal} {\bibinfo  {journal}
  {Angewandte Chemie International Edition}\ }\textbf {\bibinfo {volume}
  {42}},\ \bibinfo {pages} {4908} (\bibinfo {year} {2003})}\BibitemShut
  {NoStop}%
\bibitem [{\citenamefont {Schneider}\ \emph {et~al.}(2014)\citenamefont
  {Schneider}, \citenamefont {Matsuoka}, \citenamefont {Takeuchi},
  \citenamefont {Zhang}, \citenamefont {Horiuchi}, \citenamefont {Anpo},\ and\
  \citenamefont {Bahnemann}}]{schneider2014understanding}%
  \BibitemOpen
  \bibfield  {author} {\bibinfo {author} {\bibfnamefont {J.}~\bibnamefont
  {Schneider}}, \bibinfo {author} {\bibfnamefont {M.}~\bibnamefont {Matsuoka}},
  \bibinfo {author} {\bibfnamefont {M.}~\bibnamefont {Takeuchi}}, \bibinfo
  {author} {\bibfnamefont {J.}~\bibnamefont {Zhang}}, \bibinfo {author}
  {\bibfnamefont {Y.}~\bibnamefont {Horiuchi}}, \bibinfo {author}
  {\bibfnamefont {M.}~\bibnamefont {Anpo}}, \ and\ \bibinfo {author}
  {\bibfnamefont {D.~W.}\ \bibnamefont {Bahnemann}},\ }\href@noop {} {\bibfield
   {journal} {\bibinfo  {journal} {Chemical reviews}\ }\textbf {\bibinfo
  {volume} {114}},\ \bibinfo {pages} {9919} (\bibinfo {year}
  {2014})}\BibitemShut {NoStop}%
\bibitem [{\citenamefont {Shevlin}\ and\ \citenamefont
  {Woodley}(2010)}]{shevlin2010electronic}%
  \BibitemOpen
  \bibfield  {author} {\bibinfo {author} {\bibfnamefont {S.}~\bibnamefont
  {Shevlin}}\ and\ \bibinfo {author} {\bibfnamefont {S.}~\bibnamefont
  {Woodley}},\ }\href@noop {} {\bibfield  {journal} {\bibinfo  {journal} {The
  Journal of Physical Chemistry C}\ }\textbf {\bibinfo {volume} {114}},\
  \bibinfo {pages} {17333} (\bibinfo {year} {2010})}\BibitemShut {NoStop}%
\bibitem [{\citenamefont {Kohn}\ and\ \citenamefont {Sham}(1965)}]{Kohn-65}%
  \BibitemOpen
  \bibfield  {author} {\bibinfo {author} {\bibfnamefont {W.}~\bibnamefont
  {Kohn}}\ and\ \bibinfo {author} {\bibfnamefont {L.~J.}\ \bibnamefont
  {Sham}},\ }\href {\doibase 10.1103/PhysRev.140.A1133} {\bibfield  {journal}
  {\bibinfo  {journal} {Phys. Rev.}\ }\textbf {\bibinfo {volume} {140}},\
  \bibinfo {pages} {A1133} (\bibinfo {year} {1965})}\BibitemShut {NoStop}%
\bibitem [{\citenamefont {Himmetoglu}\ \emph {et~al.}(2014)\citenamefont
  {Himmetoglu}, \citenamefont {Floris}, \citenamefont {De~Gironcoli},\ and\
  \citenamefont {Cococcioni}}]{himmetoglu2014hubbard}%
  \BibitemOpen
  \bibfield  {author} {\bibinfo {author} {\bibfnamefont {B.}~\bibnamefont
  {Himmetoglu}}, \bibinfo {author} {\bibfnamefont {A.}~\bibnamefont {Floris}},
  \bibinfo {author} {\bibfnamefont {S.}~\bibnamefont {De~Gironcoli}}, \ and\
  \bibinfo {author} {\bibfnamefont {M.}~\bibnamefont {Cococcioni}},\
  }\href@noop {} {\bibfield  {journal} {\bibinfo  {journal} {International
  Journal of Quantum Chemistry}\ }\textbf {\bibinfo {volume} {114}},\ \bibinfo
  {pages} {14} (\bibinfo {year} {2014})}\BibitemShut {NoStop}%
\bibitem [{\citenamefont {Giannozzi}\ \emph {et~al.}(2009)\citenamefont
  {Giannozzi}, \citenamefont {Baroni}, \citenamefont {Bonini}, \citenamefont
  {Calandra}, \citenamefont {Car}, \citenamefont {Cavazzoni}, \citenamefont
  {Ceresoli}, \citenamefont {Chiarotti}, \citenamefont {Cococcioni},
  \citenamefont {Dabo}, \citenamefont {Corso}, \citenamefont {de~Gironcoli},
  \citenamefont {Fabris}, \citenamefont {Fratesi}, \citenamefont {Gebauer},
  \citenamefont {Gerstmann}, \citenamefont {Gougoussis}, \citenamefont
  {Kokalj}, \citenamefont {Lazzeri}, \citenamefont {Martin-Samos},
  \citenamefont {Marzari}, \citenamefont {Mauri}, \citenamefont {Mazzarello},
  \citenamefont {Paolini}, \citenamefont {Pasquarello}, \citenamefont
  {Paulatto}, \citenamefont {Sbraccia}, \citenamefont {Scandolo}, \citenamefont
  {Sclauzero}, \citenamefont {Seitsonen}, \citenamefont {Smogunov},
  \citenamefont {Umari},\ and\ \citenamefont {Wentzcovitch}}]{Giannozzi_2009}%
  \BibitemOpen
  \bibfield  {author} {\bibinfo {author} {\bibfnamefont {P.}~\bibnamefont
  {Giannozzi}}, \bibinfo {author} {\bibfnamefont {S.}~\bibnamefont {Baroni}},
  \bibinfo {author} {\bibfnamefont {N.}~\bibnamefont {Bonini}}, \bibinfo
  {author} {\bibfnamefont {M.}~\bibnamefont {Calandra}}, \bibinfo {author}
  {\bibfnamefont {R.}~\bibnamefont {Car}}, \bibinfo {author} {\bibfnamefont
  {C.}~\bibnamefont {Cavazzoni}}, \bibinfo {author} {\bibfnamefont
  {D.}~\bibnamefont {Ceresoli}}, \bibinfo {author} {\bibfnamefont {G.~L.}\
  \bibnamefont {Chiarotti}}, \bibinfo {author} {\bibfnamefont {M.}~\bibnamefont
  {Cococcioni}}, \bibinfo {author} {\bibfnamefont {I.}~\bibnamefont {Dabo}},
  \bibinfo {author} {\bibfnamefont {A.~D.}\ \bibnamefont {Corso}}, \bibinfo
  {author} {\bibfnamefont {S.}~\bibnamefont {de~Gironcoli}}, \bibinfo {author}
  {\bibfnamefont {S.}~\bibnamefont {Fabris}}, \bibinfo {author} {\bibfnamefont
  {G.}~\bibnamefont {Fratesi}}, \bibinfo {author} {\bibfnamefont
  {R.}~\bibnamefont {Gebauer}}, \bibinfo {author} {\bibfnamefont
  {U.}~\bibnamefont {Gerstmann}}, \bibinfo {author} {\bibfnamefont
  {C.}~\bibnamefont {Gougoussis}}, \bibinfo {author} {\bibfnamefont
  {A.}~\bibnamefont {Kokalj}}, \bibinfo {author} {\bibfnamefont
  {M.}~\bibnamefont {Lazzeri}}, \bibinfo {author} {\bibfnamefont
  {L.}~\bibnamefont {Martin-Samos}}, \bibinfo {author} {\bibfnamefont
  {N.}~\bibnamefont {Marzari}}, \bibinfo {author} {\bibfnamefont
  {F.}~\bibnamefont {Mauri}}, \bibinfo {author} {\bibfnamefont
  {R.}~\bibnamefont {Mazzarello}}, \bibinfo {author} {\bibfnamefont
  {S.}~\bibnamefont {Paolini}}, \bibinfo {author} {\bibfnamefont
  {A.}~\bibnamefont {Pasquarello}}, \bibinfo {author} {\bibfnamefont
  {L.}~\bibnamefont {Paulatto}}, \bibinfo {author} {\bibfnamefont
  {C.}~\bibnamefont {Sbraccia}}, \bibinfo {author} {\bibfnamefont
  {S.}~\bibnamefont {Scandolo}}, \bibinfo {author} {\bibfnamefont
  {G.}~\bibnamefont {Sclauzero}}, \bibinfo {author} {\bibfnamefont {A.~P.}\
  \bibnamefont {Seitsonen}}, \bibinfo {author} {\bibfnamefont {A.}~\bibnamefont
  {Smogunov}}, \bibinfo {author} {\bibfnamefont {P.}~\bibnamefont {Umari}}, \
  and\ \bibinfo {author} {\bibfnamefont {R.~M.}\ \bibnamefont {Wentzcovitch}},\
  }\href {\doibase 10.1088/0953-8984/21/39/395502} {\bibfield  {journal}
  {\bibinfo  {journal} {Journal of Physics: Condensed Matter}\ }\textbf
  {\bibinfo {volume} {21}},\ \bibinfo {pages} {395502} (\bibinfo {year}
  {2009})}\BibitemShut {NoStop}%
\bibitem [{\citenamefont {Kresse}\ and\ \citenamefont
  {Joubert}(1999)}]{PAW-PPS}%
  \BibitemOpen
  \bibfield  {author} {\bibinfo {author} {\bibfnamefont {G.}~\bibnamefont
  {Kresse}}\ and\ \bibinfo {author} {\bibfnamefont {D.}~\bibnamefont
  {Joubert}},\ }\href {\doibase 10.1103/PhysRevB.59.1758} {\bibfield  {journal}
  {\bibinfo  {journal} {Phys. Rev. B}\ }\textbf {\bibinfo {volume} {59}},\
  \bibinfo {pages} {1758} (\bibinfo {year} {1999})}\BibitemShut {NoStop}%
\bibitem [{\citenamefont {Perdew}\ \emph {et~al.}(1996)\citenamefont {Perdew},
  \citenamefont {Burke},\ and\ \citenamefont {Ernzerhof}}]{PBE-1996}%
  \BibitemOpen
  \bibfield  {author} {\bibinfo {author} {\bibfnamefont {J.~P.}\ \bibnamefont
  {Perdew}}, \bibinfo {author} {\bibfnamefont {K.}~\bibnamefont {Burke}}, \
  and\ \bibinfo {author} {\bibfnamefont {M.}~\bibnamefont {Ernzerhof}},\ }\href
  {\doibase 10.1103/PhysRevLett.77.3865} {\bibfield  {journal} {\bibinfo
  {journal} {Phys. Rev. Lett.}\ }\textbf {\bibinfo {volume} {77}},\ \bibinfo
  {pages} {3865} (\bibinfo {year} {1996})}\BibitemShut {NoStop}%
\bibitem [{\citenamefont {Kokalj}\ and\ \citenamefont
  {Caus{\`a}}(2001)}]{kokalj2001xcrysden}%
  \BibitemOpen
  \bibfield  {author} {\bibinfo {author} {\bibfnamefont {A.}~\bibnamefont
  {Kokalj}}\ and\ \bibinfo {author} {\bibfnamefont {M.}~\bibnamefont
  {Caus{\`a}}},\ }\href@noop {} {\enquote {\bibinfo {title}
  {Xcrysden:(x-window) crystalline structures and densities},}\ } (\bibinfo
  {year} {2001})\BibitemShut {NoStop}%
\bibitem [{\citenamefont {Monkhorst}\ and\ \citenamefont
  {Pack}(1976)}]{Monkhorst-76}%
  \BibitemOpen
  \bibfield  {author} {\bibinfo {author} {\bibfnamefont {H.~J.}\ \bibnamefont
  {Monkhorst}}\ and\ \bibinfo {author} {\bibfnamefont {J.~D.}\ \bibnamefont
  {Pack}},\ }\href {\doibase 10.1103/PhysRevB.13.5188} {\bibfield  {journal}
  {\bibinfo  {journal} {Phys. Rev. B}\ }\textbf {\bibinfo {volume} {13}},\
  \bibinfo {pages} {5188} (\bibinfo {year} {1976})}\BibitemShut {NoStop}%
\bibitem [{\citenamefont {Setyawan}\ and\ \citenamefont
  {Curtarolo}(2010)}]{setyawan2010high}%
  \BibitemOpen
  \bibfield  {author} {\bibinfo {author} {\bibfnamefont {W.}~\bibnamefont
  {Setyawan}}\ and\ \bibinfo {author} {\bibfnamefont {S.}~\bibnamefont
  {Curtarolo}},\ }\href@noop {} {\bibfield  {journal} {\bibinfo  {journal}
  {Computational materials science}\ }\textbf {\bibinfo {volume} {49}},\
  \bibinfo {pages} {299} (\bibinfo {year} {2010})}\BibitemShut {NoStop}%
\bibitem [{\citenamefont {Murnaghan}(1944)}]{murnaghan1944compressibility}%
  \BibitemOpen
  \bibfield  {author} {\bibinfo {author} {\bibfnamefont {F.}~\bibnamefont
  {Murnaghan}},\ }\href@noop {} {\bibfield  {journal} {\bibinfo  {journal}
  {Proceedings of the national academy of sciences of the United States of
  America}\ }\textbf {\bibinfo {volume} {30}},\ \bibinfo {pages} {244}
  (\bibinfo {year} {1944})}\BibitemShut {NoStop}%
\bibitem [{\citenamefont {Li}\ \emph {et~al.}(2018)\citenamefont {Li},
  \citenamefont {Yang}, \citenamefont {Shu}, \citenamefont {Wan}, \citenamefont
  {Wei}, \citenamefont {Yu}, \citenamefont {Breese}, \citenamefont
  {Venkatesan}, \citenamefont {Xue}, \citenamefont {Liu} \emph
  {et~al.}}]{li2018titanium}%
  \BibitemOpen
  \bibfield  {author} {\bibinfo {author} {\bibfnamefont {Y.}~\bibnamefont
  {Li}}, \bibinfo {author} {\bibfnamefont {Y.}~\bibnamefont {Yang}}, \bibinfo
  {author} {\bibfnamefont {X.}~\bibnamefont {Shu}}, \bibinfo {author}
  {\bibfnamefont {D.}~\bibnamefont {Wan}}, \bibinfo {author} {\bibfnamefont
  {N.}~\bibnamefont {Wei}}, \bibinfo {author} {\bibfnamefont {X.}~\bibnamefont
  {Yu}}, \bibinfo {author} {\bibfnamefont {M.~B.}\ \bibnamefont {Breese}},
  \bibinfo {author} {\bibfnamefont {T.}~\bibnamefont {Venkatesan}}, \bibinfo
  {author} {\bibfnamefont {J.~M.}\ \bibnamefont {Xue}}, \bibinfo {author}
  {\bibfnamefont {Y.}~\bibnamefont {Liu}},  \emph {et~al.},\ }\href@noop {}
  {\bibfield  {journal} {\bibinfo  {journal} {chemistry of materials}\ }\textbf
  {\bibinfo {volume} {30}},\ \bibinfo {pages} {4383} (\bibinfo {year}
  {2018})}\BibitemShut {NoStop}%
\bibitem [{\citenamefont {Niu}\ \emph {et~al.}(2011)\citenamefont {Niu},
  \citenamefont {Xu}, \citenamefont {Shao},\ and\ \citenamefont
  {Cheng}}]{niu2011enhanced}%
  \BibitemOpen
  \bibfield  {author} {\bibinfo {author} {\bibfnamefont {M.}~\bibnamefont
  {Niu}}, \bibinfo {author} {\bibfnamefont {W.}~\bibnamefont {Xu}}, \bibinfo
  {author} {\bibfnamefont {X.}~\bibnamefont {Shao}}, \ and\ \bibinfo {author}
  {\bibfnamefont {D.}~\bibnamefont {Cheng}},\ }\href@noop {} {\bibfield
  {journal} {\bibinfo  {journal} {Applied Physics Letters}\ }\textbf {\bibinfo
  {volume} {99}},\ \bibinfo {pages} {203111} (\bibinfo {year}
  {2011})}\BibitemShut {NoStop}%
\bibitem [{\citenamefont {Persson}\ and\ \citenamefont
  {Mirbt}(2006)}]{persson2006improved}%
  \BibitemOpen
  \bibfield  {author} {\bibinfo {author} {\bibfnamefont {C.}~\bibnamefont
  {Persson}}\ and\ \bibinfo {author} {\bibfnamefont {S.}~\bibnamefont
  {Mirbt}},\ }\href@noop {} {\bibfield  {journal} {\bibinfo  {journal}
  {Brazilian journal of physics}\ }\textbf {\bibinfo {volume} {36}},\ \bibinfo
  {pages} {286} (\bibinfo {year} {2006})}\BibitemShut {NoStop}%
\bibitem [{\citenamefont {Hu}\ and\ \citenamefont
  {Metiu}(2011)}]{hu2011choice}%
  \BibitemOpen
  \bibfield  {author} {\bibinfo {author} {\bibfnamefont {Z.}~\bibnamefont
  {Hu}}\ and\ \bibinfo {author} {\bibfnamefont {H.}~\bibnamefont {Metiu}},\
  }\href@noop {} {\bibfield  {journal} {\bibinfo  {journal} {The Journal of
  Physical Chemistry C}\ }\textbf {\bibinfo {volume} {115}},\ \bibinfo {pages}
  {5841} (\bibinfo {year} {2011})}\BibitemShut {NoStop}%
\end{thebibliography}%
\end{document}